# Stepwise Refinement of Data Flow Architectures *

Jan Philipps      Bernhard Rumpe

Institut für Informatik
Technische Universität München
D-80290 München
{philipps,rumpe}@informatik.tu-muenchen.de

## Abstract

*Software and hardware architectures are prone to modifications. We demonstrate how a mathematically founded refinement calculus for a class of architectures, namely data flow networks, can be used to modify a system in a provably correct way. The calculus consists of basic rules to add and to remove components and channels to a system.*

## 1 Introduction

The architecture of a software or hardware system influences its efficiency, its adaptility, and the reusability of components. Especially the adaption to new requirements causes frequent changes in the architecture while the system is developed, or when it is later extended. However, the definition of architecture is still rather informal in the software engineering community, and the question of how to properly modify an architecture has not been adequately addressed so far.

In this paper, we examine how a certain class of system architectures, namely data flow networks, can be modified, so that the new system is a provably correct refinement of the original system. Our work is based on a precise mathematical model [2, 3, 4] for such data flow networks. This model gives a compositional semantics to data flow networks, and hence components can be structurally composed to build hierarchical models of a system.

The semantic model is simple, yet powerful: when specifying component behavior, certain aspects can be left open. We refer to this style as *underspecification*. The reduction

* This paper partly originates from the SysLab project, which is supported by the DFG under the Leibniz program, by Siemens-Nixdorf and Siemens Corporate Research.

of this underspecification immediately gives a refinement relation for black box behaviors.

In addition to black-box or *behavioral* refinement, two other classes of refinement relation can be established:

- structural refinement (glass box refinement)
- signature refinement

While black box refinement only relates black box behaviors of not further detailed components, structural refinement allows us to refine a black box behavior by a subsystem architecture. Signature refinement deals with the manipulation of the system or component interfaces. As shown in [1], both structural and signature refinement can be reduced to behavioral refinement. In Section 2 we will see that behavioral refinement is a simple subset relation.

Neither of these three refinement classes, however, allows architectural refinement in the sense that two glassbox architectures are related. In [10], we introduced a concept for glass-box refinement; again, it can be defined in terms of behavioral refinement. For the practical application of architectural refinement, we defined a rule system to incrementally change an architecture, e.g. by adding new components or channels.

In this paper, we demonstrate in detail how the rule system can be applied to a concrete example. It is structured as follows. In Sections 2 and 3 we present the mathematical foundations and define the concepts of component and system. In Section 4 we summarize the rules introduced in [10]. Section 5 describes the refinement of the refinement of a simple data acquisition system. Section 6 concludes.

## 2 Semantic Model

In this section we introduce the basic mathematical concepts for the description of systems. We concentrate on interactive systems that communicate asynchronously through channels. A component is modeled as a relation over input and output communication histories that obeys certain causality constraints.

We assume that there is a given set of channel identifiers, $\mathbb{C}$, and a given set of messages, $M$.

**Streams.** We use *streams* to describe communication histories on channels. A stream over the set $M$ is a finite or infinite sequence of elements from $M$. By $M^*$ we denote the finite sequences over the set $M$. The set $M^*$ includes the empty sequence that we write as $\langle\,\rangle$. The set of infinite sequences over $M$ is denoted by $M^\infty$.

Communication histories are represented by *timed streams*:

$$M^{\aleph} =_{def} (M^*)^\infty$$

The intuition is that the time axis is divided into an infinite stream of time intervals, where in each interval a finite number of messages may be transmitted.

These intervals are often of a fixed duration, such as months or days for reports in business information systems, or milliseconds in more technical applications. Their duration need not be fixed, however: the intervals could also span the time between certain events that are of interest to the system, such as the pressing of a button. In each interval, the order of the messages is fixed, but the exact arrival time of a message is unknown.

For $i \in \mathbb{N}$ and $x \in M^{\mathbb{N}}$ we denote by $x \downarrow i$ the sequence of the first $i$ sequences in the stream $x$. When writing specifications, we sometimes ignore the interval boundaries, and regard a stream as the finite or infinite sequence of messages that results from the concatenation of all the intervals. We then use the syntax $a \,\&\, r$ to split a stream into its first element $a$, and the remaining sequence $r$.

A *named stream tuple* is a function $\mathbb{C} \to M^{\mathbb{N}}$ that assigns histories to channel names. For $C \subseteq \mathbb{C}$ we write $\overrightarrow{C}$ for the set of named stream tuples with domain $C$.

For $x \in \overrightarrow{C}$ and $C' \subseteq C$, the named stream tuple $x \mid_{C'} \in \overrightarrow{C'}$ denotes the restriction of $x$ to the channels in $C'$:

$$\forall\, c \in C' : x \mid_{C'} (c) = x(c)$$

**Behaviors.** We model the interface behavior of a component with the set of input channels $I \subseteq \mathbb{C}$ and the set of output channels $O \subseteq \mathbb{C}$ by a function

$$\beta : \overrightarrow{I} \to \mathbb{P}(\overrightarrow{O})$$

Intuitively, $\beta$ maps the incoming input on $I$ to the set of possible outputs on $O$, and thus describes the visible behavior of a component with input channels $I$ and outputs channels $O$.

Equivalently, $\beta$ can be seen as a relation over the named stream tuples in $\overrightarrow{I}$ and the named stream tuples in $\overrightarrow{O}$. $\beta$ is called a *behavior*. Since for every input history multiple output histories can be allowed by a behavior, it is possible to model nondeterminism, or equivalently, to regard relations with multiple outputs for one input as underspecified.

A function $f \in \overrightarrow{I} \to \overrightarrow{O}$ can be seen as a special case of a deterministic relation. We say the $f$ is *time guarded*, iff for all input histories $x$ and $y$, and for all $i \in \mathbb{N}$

$$x \downarrow i = y \downarrow i \Rightarrow (f\ x) \downarrow (i+1) = (f\ y) \downarrow (i+1)$$

A time guarded function $f$ is called a *strategy* for a behavior $\beta$ if for all $x$ we have $f(x) \in \beta(x)$. If $\beta$ has at least one strategy, we say that $\beta$ is *realizable*.

Time guardedness reflects the notion of time and causality. The output at a certain time interval may only depend on the input received so far, and not on future input.

**Interface adaption.** Given a behavior $\beta : \overrightarrow{I} \to \mathbb{P}(\overrightarrow{O})$, we can define a behavior with a different interface by extending the set of input channels, and restricting the set of output channels. If $I \subseteq I'$ and $O' \subseteq O$, then $\beta' = \beta \updownarrow_{O'}^{I'}$ is again a behavior with $\beta'(i) = (\beta(i \mid_I)) \mid_O$.

This corresponds to the change of the component interface by adding input channels that are ignored by the component, and by removing output channels that are ignored by the environment.

**Composition.** Behaviors can be composed by a variety of operators. Sequential and parallel composition, as well as a feedback construction is introduced in [6]. For our work, we use a generalized operator $\otimes$ that composes a finite set of behaviors

$$B = \{\beta_1 : \overrightarrow{I_1} \to \mathbb{P}(\overrightarrow{O_1}), \ldots, \beta_n : \overrightarrow{I_n} \to \mathbb{P}(\overrightarrow{O_n})\}$$

in parallel with implicit feedback. We define

$$O = \cup_{1 \leq k \leq n} O_k$$
$$I = (\cup_{1 \leq k \leq n} I_k) \setminus O$$

where $O$ is the union of all component outputs, and $I$ is the set of those inputs, that are not connected to any of the components' outputs.

Then the relation $\otimes B \in \overrightarrow{I} \to \mathbb{P}(\overrightarrow{O})$ is characterized by:

$$o \in (\otimes B)(i) \Leftrightarrow$$
$$\exists\, l \in \overrightarrow{(I \cup O)} :$$
$$l\mid_O = o \wedge l\mid_I = i \wedge$$
$$\forall\, k \in \{1, \ldots n\} : l\mid_{O_k} \in \beta_k(l\mid_{I_k})$$

If all behaviors in $B$ are realizable, then so is $\otimes B$. The proof follows [6]; it relies on the time guardedness of strategy functions.

It is easy to express parallel and sequential composition of behaviors with the $\otimes$ operator.

**Refinement.** Intuitively, a behavior describes the externally observable input/output relation that the clients of a component may rely on. Refining a behavior in a modular way means that the client's demands are still met, when the component behavior is specialized.

Formally, the refinement relation in our framework is defined as follows. Given two behaviors $\beta_1, \beta_2 \in \overrightarrow{I} \to \mathbb{P}(\overrightarrow{O})$ we say that $\beta_1$ is refined by $\beta_2$, iff

$$\forall\, i \in \overrightarrow{I} : \beta_2(i) \subseteq \beta_1(i)$$

Refinement means in our context that each possible channel history of the new component is also a possible channel history of the original component.

## 3 Components and Systems

In this section, we define an abstract notion of system architecture. Basically, a system consists of a set of *components* and their *connections*. We first define components, and then introduce the architectural or glass box view, and the black box view of a system.

**Components.** A *component* is a tuple $c = (n, I, O, \beta)$, where $n$ is the name of the component, $I \subseteq \mathbb{C}$ is the set of input channels, and $O \subseteq \mathbb{C}$ the set of output channels. Moreover, $\beta : \overrightarrow{I} \to \mathbb{P}(\overrightarrow{O})$ is a behavior.

The operators name.$c$, in.$c$, out.$c$ and behav.$c$ yield $n$, $I$, $O$ and $\beta$, respectively. The name $n$ is introduced mainly as a convenience for the system designer. The channel identifiers in.$c$ and out.$c$ define the interface of the component.

**Architectural view of a system.** In the architectural view, a system comprises a finite set of components. A connection between components is established by using the same channel name.

A system is thus a tuple $S = (I, O, C)$, where $I \subseteq \mathbb{C}$ is the input interface, and $O \subseteq \mathbb{C}$ is the output interface of the system. $C$ is a finite set of components.

We want to be able to decompose systems hierarchically. In fact, as we will see, a system can be regarded as an ordinary component. Therefore systems need not be closed (having empty interfaces), and we introduce the interface channels $I$ and $O$ to distinguish external from internal channels.

We define the operators in.$S$, out.$S$, arch.$S$ to return $I$, $O$ and $C$, respectively. In addition, we write:

$$\text{in}.C =_{def} \cup_{c \in \text{arch}.S}(\text{in}.c)$$
$$\text{out}.C =_{def} \cup_{c \in \text{arch}.S}(\text{out}.c)$$

for the union of the input or output interfaces, respectively, of the components of $S$.

The following consistency conditions ensure a meaningful architectural view of a system $S$. Let $c, c_1, c_2 \in \text{arch}.S$ be components, with $c_1 \neq c_2$.

| | | | |
|---|---|---|---|
| (1) | $\text{name}.c_1 \neq \text{name}.c_2$ | | *Different components have different names* |
| (2) | $\text{out}.c_1 \cap \text{out}.c_2 = \emptyset$ | | *Each channel is controlled by only one component* |
| (3) | $\text{in}.S \cap \text{out}.c = \emptyset$ | | *Input channels of the system interface are controlled by the environment, not by a component* |
| (4) | $\text{in}.c \subseteq \text{out}.C \cup \text{in}.S$ | | *Each input channel of a component controlled by either another component or by the environment* |
| (5) | $\text{out}.S \subseteq \text{out}.C$ | | *Each channel of the output interface is controlled by a component* |

Note that we allow that input channels are in more than one interface: a channel can have multiple readers, even broadcasting is possible. Not every channel of the system input interface has to be connected to a component, since condition 4 only requires the subset relation instead of equality.

We allow a component to read and write on the same channel if desired; as a consequence of conditions (3) and (5), however, system input and output are disjoint.

**Black Box view of a system.** The behavior of a component $c$ is given in terms of its relation behav.$c$ between input and output streams. We define the *black box behavior* of a system $S$ composed of finitely many components arch.$S$ using the composition operator $\otimes$. The result of this composition is then made compatible with the system interface by restricting the output channels to those in out.$S$, and by extending the input channels to those in in.$S$:

$$[\![S]\!] = (\otimes\{\text{ behav}.c \mid c \in \text{arch}.S \})\updownarrow_{\text{out}.S}^{\text{in}.S}$$

Because of the context conditions for systems the composition is well-defined. The hiding of the internal output channels $\text{out}.C \setminus \text{out}.S$ and the extension with the unused input channels $\text{in}.S \setminus \text{in}.C$ is also well-defined.

The black box behavior has the signature:

$$[\![S]\!] : \overrightarrow{\text{in}.S} \to \mathbb{P}(\overrightarrow{\text{out}.S})$$

Thus, the black box behavior can now be used as a component description itself. Introducing a fresh name $n$, we define the component $c_S$ as:

$$c_S = (n, \text{in}.S, \text{out}.S, [\![S]\!])$$

In this way, a hierachy of architectural views can be defined and iteratively refined and detailed.

Later on we need a more detailed definition of this semantics. By expanding the definitions of the $\otimes$ and $\updownarrow$ operators, we obtain the following equivalent characterisation of $[\![(I, O, C)]\!]$:

$$\begin{aligned}
o \in [\![(I, O, C)]\!](i) &\Leftrightarrow \\
\exists\, l \in &\overrightarrow{(I \cup \text{out}.C)} : \\
&l\mid_O = o \wedge l\mid_I = i \wedge \\
&\quad \forall\, c \in C :\ l\mid_{\text{out}.c} \in (\text{behav}.c)(l\mid_{\text{in}.c})
\end{aligned}$$

This expanded characterisation says, that $o$ is an output of the system for input $i$ (line 1), iff there is a mapping $l$ of all channels to streams (line 2), such that $l$ coincides with the given input $i$ and output $o$ on the system interface channels (line 3) and feeding the proper submapping of $l$ into a component results also in a submapping of $l$.

# 4 Refinement of system architectures

When a system is refined, it must not break the interaction with its environment. The observable behavior of a refined system must be a refinement of the behavior of the original system.

In this paper, we leave the interface of the system unchanged. Interface refinements that affect the signature of a system $S$ are described in [1] for black box behaviors; they can be adapted to our architectural framework. We also ignore aspects of realizability. The techniques used to prove that a component specification is realizable are orthogonal to the rules presented here, and will not be considered in this paper.

We therefore define the refinement relation on systems as a behavioral refinement on the given interface:

$$S \rightsquigarrow S' \Leftrightarrow_{def} \forall\, i \in \overrightarrow{\text{in}.S} : [\![S']\!](i) \subseteq [\![S]\!](i)$$

As explained above, we tacitly assume that $\text{in}.S = \text{in}.S'$ and $\text{out}.S = \text{out}.S'$. Stepwise refinement is possible, since the refinement relation is transitive:

$$S \rightsquigarrow S' \wedge S' \rightsquigarrow S'' \Rightarrow S \rightsquigarrow S''$$

In [10], we defined and justified a set of constructive refinement rules that allows refinements of system architectures. The rules allow us to add and remove components, to add and remove channels, to refine the behavior of components, and to refine single components to subsystems and vice versa.

In the sequel, we summarize these rules; in Section 5 we will apply them to a simple example. Each rule refines a system $S = (I, O, C)$ into another system $S' = (I, O, C')$. We use the syntax

$$S \text{ WITH } C := C'$$

to denote the system $(I, O, C')$. In addition, we write

$$S \text{ WITH } c := c'$$

to denote the system $(I, O, (C \setminus \{c\}) \cup \{c'\})$.

To create a component with the same name and interface as $c = (n, I, O, \beta)$, but with a different behavior $\beta'$, we use the syntax

$$c \text{ WITH } \mathsf{behav}.c := \beta'$$

to denote the component $(n, I, O, \beta')$. Similarly, we can change the name or interface of a component.

The refinement rules are presented in the syntax

$$\begin{array}{|c} (Premises) \\ \hline (Refinement) \end{array}$$

where the premises are conditions to be fulfilled for the refinement relation to hold.

**Behavioral refinement.** Systems can be refined by refining the behavior of their components. Let $c \in C$ be a component. If we refine the behavior of $c$ to $\beta$, we get a refinement of the externally visible, global system behavior:

$$\begin{array}{|c} c \in C \\ \forall\, i \in \overrightarrow{\mathsf{in}.c} : \beta(i) \subseteq \mathsf{behav}.c \\ \hline S \leadsto S \text{ WITH } \mathsf{behav}.c := \beta \end{array}$$

In some cases, to prove the behavioral refinement of $c$ some assumptions on the contents of $c$'s input channels are necessary. Then this simple rule cannot be used.

To overcome this problem, we introduce the notion of behavioral refinement in the context of an *invariant*. An invariant is a predicate $\Psi$ over the possible message flows within a system $S = (I, O, C)$:

$$\Psi : \overrightarrow{(I \cup \mathsf{out}.C)} \to \mathbb{B}$$

An invariant is valid within a system, if it holds for all named stream tuples $l$ defining the system's streams. This can be formally expressed similar to the expanded definition of the system semantics $[\![S]\!]$ presented in Section 3:

$$\forall\, l \in \overrightarrow{(I \cup \mathsf{out}.C)} :$$
$$\quad (\forall\, c \in C : l \mid_{\mathsf{out}.c} \in (\mathsf{behav}.c)(l \mid_{\mathsf{in}.c})) \Rightarrow \Psi(l)$$

Note that invariants are not allowed to restrict the possible inputs on channels from $I$, but only characterize the internal message flow.

Let us now assume that we want to replace the behavior of component $c$ by a new behavior $\beta$. The latter is a refinement of behav.$c$ under the invariant $\Psi$, when:

$$\forall\, l \in \overrightarrow{(I \cup \mathsf{out}.C)} :$$
$$\Psi(l) \Rightarrow \beta(l\mid_{\mathsf{in}.c}) \subseteq (\mathsf{behav}.c)(l\mid_{\mathsf{in}.c})$$

Thus, the complete refinement rule is as follows. The two premises express that $\Psi$ is a valid invariant, and that $\beta$ refines behav.$c$ under this invariant.

$$\dfrac{\begin{array}{l} \forall\, l \in \overrightarrow{(I \cup \mathsf{out}.C)} : \\ \quad (\forall\, c \in C : l\mid_{\mathsf{out}.c} \in (\mathsf{behav}.c)(l\mid_{\mathsf{in}.c})) \Rightarrow \Psi(l) \\ \forall\, l \in \overrightarrow{(I \cup \mathsf{out}.C)} : \\ \quad \Psi(l) \Rightarrow \beta(l\mid_{\mathsf{in}.c}) \subseteq (\mathsf{behav}.c)(l\mid_{\mathsf{in}.c}) \end{array}}{S \rightsquigarrow S \ \textsc{with}\ \mathsf{behav}.c := \beta}$$

This rule is the only one that requires global properties of a system as a premise. The other rules only deal locally with one affected component. However, since $\Psi$ is used only for a single application of this rule, it is often sufficient to prove its invariance with respect to a relevant subset of all the system components.

Behavioral refinement of a component usually leads to true behavioral refinement of the system. This is in general not the case for the following architectural refinements, which leave the global system behavior unchanged.

**Adding and removing output channels.** If a channel is neither connected to a system component, nor part of the system interface, it may be added as a new output channel to a component $c \in \mathsf{arch}.S$:

$$\dfrac{\begin{array}{l} p \in \mathbb{C} \setminus (I \cup \mathsf{out}.C) \\ \beta \in \overrightarrow{\mathsf{in}.c} \rightarrow \mathbb{P}(\overrightarrow{\mathsf{out}.c \cup \{p\}}) \\ \forall\, i, o :\ o \in \beta(i) \Leftrightarrow o\mid_{\mathsf{out}.c} \in \mathsf{behav}.c(i) \end{array}}{\begin{array}{l} S \rightsquigarrow S\ \textsc{with} \\ \qquad \mathsf{out}.c := \mathsf{out}.c \cup \{p\} \\ \qquad \mathsf{behav}.c := \beta \end{array}}$$

The new behavior $\beta$ does not restrict the possible output on channel $p$. Hence, the introduction of new output channels increases the nondeterminism of the component. It does not, however, affect the behavior of the composed system, since $p$ is neither part of the system interface nor connected to any other component. The contents of the new channel can be restricted with the behavioral refinement rule.

Similarly, an output channel $p \in \mathsf{out}.c$ can be removed from the component $c$, provided that it is not used elsewhere in the system:

$$\dfrac{\begin{array}{l} p \notin O \cup \mathsf{in}.C \\ \beta = \mathsf{behav}.c\updownarrow^{\mathsf{in}.c}_{\mathsf{out}.c \setminus \{p\}} \end{array}}{\begin{array}{l} S \rightsquigarrow S\ \textsc{with} \\ \qquad \mathsf{out}.c := \mathsf{out}.c \setminus \{p\} \\ \qquad \mathsf{behav}.c := \beta \end{array}}$$

The new behavior $\beta$ is the restriction of the component behavior behav.$c$ to the remaining channels.

Adding and removing output channels are complementary transformations. Consequently, both rules are behavior preserving. This is not surprising, since the channel in question so far is not used by any other component.

**Adding and removing input channels.** An input channel $p \in \mathbb{C}$ may be added to a component $c \in C$, if it is already connected to the output of some other component or to the input from the environment:

$$\begin{array}{|l}
p \in I \cup \mathsf{out}.C \\
\beta = \mathsf{behav}.c \updownarrow_{\mathsf{out}.c}^{\mathsf{in}.c \cup \{p\}} \\
\hline
S \leadsto S \text{ WITH} \\
\quad \mathsf{in}.c := \mathsf{in}.c \cup \{p\} \\
\quad \mathsf{behav}.c := \beta
\end{array}$$

The new behavior $\beta$ now receives input from the new input channel $p$, but is still independent of the data in $p$.

If the behavior of a component $c$ does not depend on the input from a channel $p$, the channel may be removed:

$$\begin{array}{|l}
\forall\, i, i' \in \overrightarrow{\mathsf{in}.c} : \ i \mid_{\mathsf{in}.c \setminus \{p\}} = i' \mid_{\mathsf{in}.c \setminus \{p\}} \\
\quad \Rightarrow \mathsf{behav}.c(i) = \mathsf{behav}.c(i') \\
\forall\, i \in \overrightarrow{\mathsf{in}.c} : \ \beta(i \mid_{\mathsf{in}.c \setminus \{p\}}) = \mathsf{behav}.c(i) \\
\hline
S \leadsto S \text{ WITH} \\
\quad \mathsf{in}.c := \mathsf{in}.c \setminus \{p\} \\
\quad \mathsf{behav}.c := \beta
\end{array}$$

Because the component does not depend on the input from $p$ (first premise), there is a behavior $\beta$ satisfying the second premise.

The rule for removing input channels might seem useless — why should a component have an input it does not rely on? However, note that it is possible to first add new input channels that provide basically the same information as an existing channel, then to change the component's behavior so that it relies on the new channels instead. Finally, the old channel can safely be reduced.

As with output channels, adding and removing input channels are complementary transformations and thus behavior preserving. This is because the input channels do not influence the component's behavior, and therefore the global system behavior is unchanged, too.

**Adding and removing components.** A component can be added without without changing the global system behavior if we ensure that it is not connected to the other components, or to the system environment. Later, we may successively add input or output channels, and refine the new component's behavior with the previously given rules.

$$\frac{\forall\, c \in C \,:\, \mathsf{name}.c \neq n}{S \rightsquigarrow S \text{ WITH } C := C \cup \{(n, \varnothing, \varnothing, \alpha)\}}$$

The premise simply ensures that the name $n$ is fresh; the new behavior $\alpha$ is somewhat subtle: it is the unique behavior of a component with no input and no output channels: $\{()\} = \alpha(())$.

Similarly, components may be removed if they have no output ports that might influence the functionality of the system.

$$\frac{\mathsf{out}.c = \varnothing}{S \rightsquigarrow S \text{ WITH } C := C \setminus \{c\}}$$

**Expanding and Folding.** As we have seen, components can be defined with the black box view of systems. In this way system architectures can be decomposed hierarchically: a single component of a system is replaced by another system. We therefore need a rule for expansion of components. Assume a given system architecture $S = (I_S, O_S, C_S)$ contains a component $c \in C_S$. This component $c$ is itself described by an architecture $T = (I_T, O_T, C_T)$. The names of the components in $T$ are assumed to be disjoint from those in $S$; through renaming this can always be ensured. The expansion of $T$ in $S$ takes the components and channels of $T$ and incorporates them within $S$.

$$\frac{\begin{array}{l} c = (n, I_T, O_T, [\![T]\!]) \\ \mathsf{out}.C_T \cap \mathsf{out}.C_S = \mathsf{out}.c \\ \mathsf{out}.C_T \cap I_S = \varnothing \end{array}}{S \rightsquigarrow S \text{ WITH } C_S := C_S \setminus \{c\} \cup C_T}$$

The first premise means that the architecture $T$ describes the component $c$. The other two premises require that the internal channels of $T$, which are given by $\mathsf{out}.C_T \setminus O_T$, are not used in $S$. In general, this can be accomplished through a renaming rule, which it would be straightforward to define.

The complementary operation to the expansion of a component is the folding of a subarchitecture $T = (I_T, O_T, C_T)$ of a given system $S = (I, O, C)$.

$T$ is a subarchitecture of $S$, if

- the components $C_T$ are a subset of the components $C$ of $S$;

- the inputs $I_T$ of $T$ at least include the inputs of the components in $C_T$ that are not connected to some output of a component in $C_T$; they may include other inputs as well, except those input channels that are either in the global system input $I$ or controlled by a component in the complete system $C$;

- similarly, the outputs $O_T$ are a subset of the component outputs $\mathsf{out}.C_T$, and include at least those outputs from $\mathsf{out}.C_T$ that are connected to either the environment or to other components in $C$.

The folding rule is defined as follows:

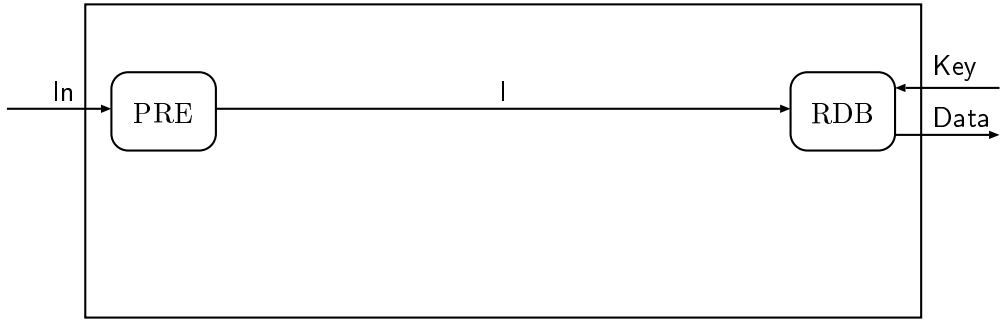

Figure 1: Database example

$$\begin{array}{|l} C_T \subseteq C \\ \mathsf{in}.C_T \setminus \mathsf{out}.C_T \ \subseteq\ I_T \ \subseteq\ (I \cup \mathsf{out}.C) \setminus O_T \\ \mathsf{out}.C_T \cap (O \cup \mathsf{in}.(C \setminus C_T)) \subseteq O_T \subseteq \mathsf{out}.C_T \\ \forall\, c \in C \setminus C_T :\ \mathsf{name}.c \neq n \\ \hline S \rightsquigarrow S \ \textsc{with}\ C := C \setminus C_T \cup \{(n, I_T, O_T, [\![T]\!])\} \end{array}$$

The first three premises are the conditions mentioned above; the fourth premise requires that the name $n$ of the new component is not used elsewhere in the resulting system.

## 5 Refinement example

In this section, we demonstrate how our refinement rule system can be used in practice. Our example architecture is shown in Figure 1; it models a small data acquisition system.

The system reads input via an input In; the messages on In consist of pairs of a key and some data to be stored under this key; new data values for the same key overwrite old values. Concurrently, the system answers request for the data of a certain key that is input via channel Key by transmitting the data stored in the database under this key via channel Data.

Internally, the system consists of two components: a preprocessor PRE, and a database RDB. The data from the environment first undergoes some transformations in PRE, and is then forwarded via the internal channel I to the remote database.

Let $Key$ be the set of possible keys for the database, and $Data$ the set of possible data values. Then, $Entry = Key \times Data$ is the set of possible entries for the database. The database itself is modeled as a function $M : Key \to Data$. We write $M(k)$ for the data item stored under key $k$. If there is not yet a proper item stored under $k$, then $M(k)$ should return an otherwise unused item $\bot$. By $M[k \mapsto d]$ we denote the updated database $M'$, where $M'(j) = d$ if $j = k$, and $M'(j) = M(j)$ otherwise.

The two components PRE and RDB are specified as state machines (Figures 2, 3). We assume that there is a given function $f : Data \to Data$, that handles the preprocessing for a single datum.

In order to reduce the transmission time for the entries, we now want to transmit for each entry only the difference of the entry's data with respect to the already stored data for that key; the differences are assumed to be smaller in size than the data itself. Of course, the first entry for each key will need to be transmitted completely.

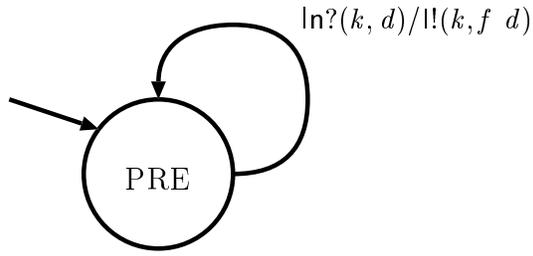

Figure 2: Preprocessor specification

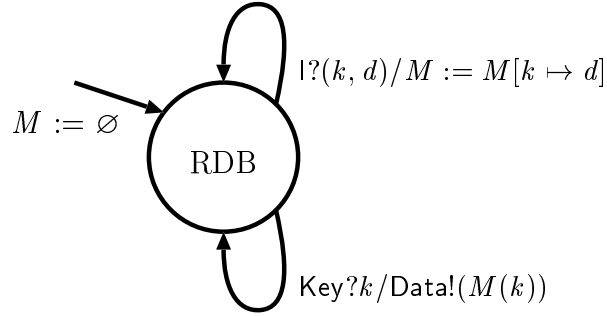

Figure 3: Remote database specification

We are not intersted in the algorithmic aspects of the computation of the difference between old data and new data; we just assume that the difference between two data items is itself an element of *Data*, and that there is a function

$$\Delta : Data \times Data \to Data$$

that computes the difference between old and new data. Another function

$$\rho : Data \times Data \to Delta$$

reconstructs the new data given old data and the difference. We require that

$$\rho(d_{old}, \Delta(d_{old}, d_{new})) = d_{new}$$

To simplify our specifications, we also assume that

$$\Delta(\bot, d) = d, \quad \rho(\bot, \delta) = \delta$$

These two function can be extended to streams, where they take a database M as an additional parameter:

$$\Delta^*_M (\langle \rangle) = \langle \rangle$$
$$\Delta^*_M ((k, d) \& x) = (k, \Delta(M(k), d)) \& \Delta^*_{M[k \mapsto d]} (x)$$

$$\rho^*_M (\langle \rangle) = \langle \rangle$$
$$\rho^*_M ((k, \delta) \& x) = (k, \rho(M(k), \delta)) \& \rho^*_{M[k \mapsto \rho(M(k),\delta)]} (x)$$

Informally, the system modification is simple: the preprocessor is extended with a local database; for each new entry the difference to the old is computed and forwarded. The remote database reads the input, computes the new value out of stored value and received difference, and stores this new value in its database. One possible design for this modification is to introduce encoding and decoding components, that compute the differences and reconstruct the original data, respectively.

In the sequel, we show how this refinement can be justified with our rule system. The modification consists of eight steps.

**Step 1: Adding components.** First, we introduce two new components to the system by two applications of the refinement rule. The new components, ENC and DEC, are not connected to any other component in the system.

After this refinement step, the system looks as follows:

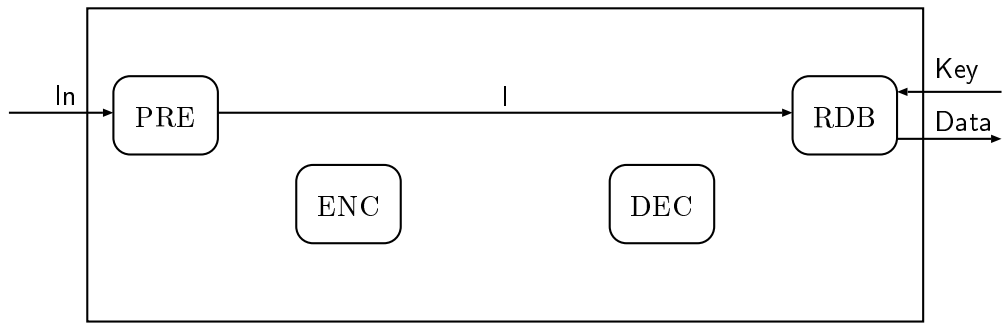

**Step 2: Adding output channels.** Now we add an output channel D to ENC, and an output channel R to DEC. Since these channels are neither part of the system interface, nor previously connected to any component, the premises of the refinement rule for the addition of channels are satisfied. Note that the contents of the channel are so far completely undefined, and the components ENC and DEC are therefore now nondeterministic. Nevertheless, the behavior of the system itself is unchanged, since the data on the new channels is unused throughout the system.

The following figure depicts the system after this refinement step;

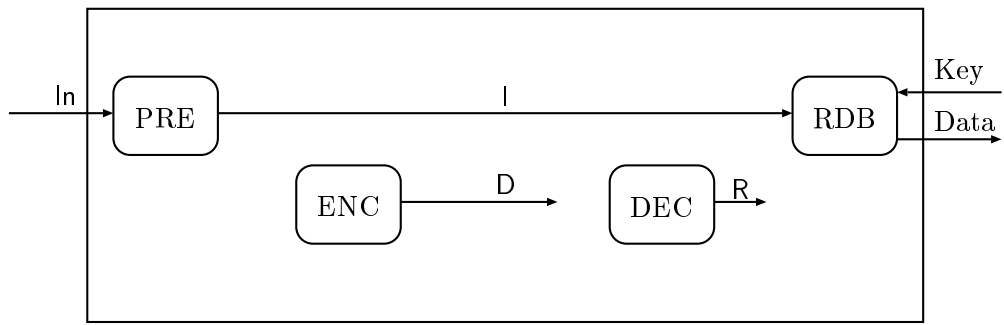

**Step 3: Adding input channels.** We now connect the channel I to the encoder ENC. The encoder still ignores the additional input, however, and hence the output D of ENC is still arbitrary. Similarly, we connect D to the decoder.

The system now looks like this:

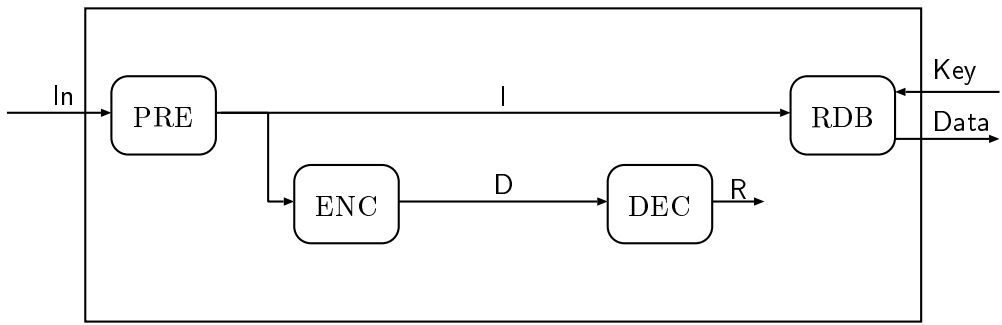

**Step 4: Behavioral refinement.** Now we constrain the channels D and R to carry the differences of the data on I and the reconstructed data, respectively. This is accomplished by restricting the behavior of ENC and DEC, and we can use the simple behavioral refinement rule for this step.

The encoder component is now specified as follows:

$$(\text{ENC}, \{\mathsf{I}\}, \{\mathsf{D}\}, \beta)$$

where

$$\forall\, l, l' : l' \in \beta(l) \Leftrightarrow l'(\mathsf{D}) = \Delta^*_\varnothing(l(\mathsf{I}))$$

Thus, the encoder just applies the difference function $\Delta^*$ to its input stream I.

Similarly, we define the behavior of DEC as an application of the restoration function $\rho$. Since until now the behavior of the components was completely unspecified, this refinement is obviously correct.

The structure of the system remains unchanged.

**Step 5: Adding an input channel.** We now connect the channel R to the remote database. The behavior of RDB still ignores the additional input, however.

This step gives us the following system:

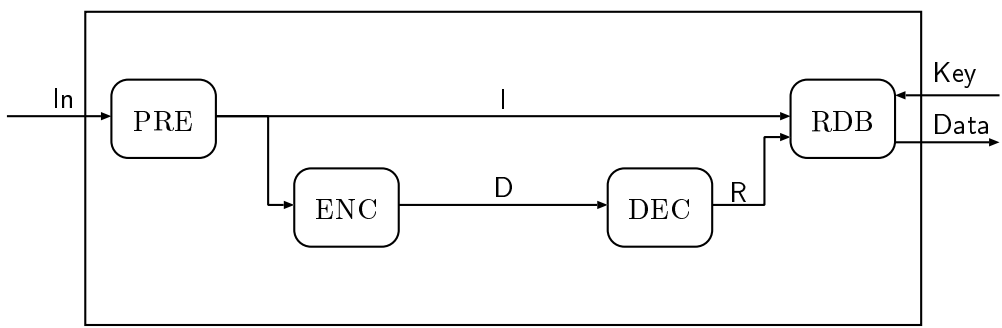

**Step 6: Behavioral refinement with invariant.** Now we want the remote database to store the data transmitted on R instead of that on I. Conversely, the input via I should be ignored.

The new behavior can again be specified as a state transition diagram; it looks just like the one in Figure 3, except that the upper transition reads from channel R instead of channel I.

Unfortunately, we cannot prove this refinement step with the simple behavioral refinement rule used in Step 4. The reason is that after the refinement the behavior of RDB is only then still correct, if the data on R is the same as that on I. Since neither R nor I is controlled by RDB, this cannot be proven locally.

The solution here is to use the behavioral refinement rule with an invariant. Intuitively, we know that encoding and then decoding the processed data from PRE yields the same data as that on I.

We can formalize this knowledge with the following invariant:

$$\Psi(l) =_{def} l(\mathsf{I}) = \rho^*_\varnothing(\Delta^*_\varnothing(l(\mathsf{I})))$$

To show that $\Psi$ is indeed an invariant we prove the following property, which implies $\Psi$:

$$\forall\, x, \forall\, M : \rho^*_M(\Delta^*_M(x)) = x$$

The proof is by induction on $x$:

- If $x = \langle\,\rangle$, we have for all $M$: $\Delta^*_M(x) = \langle\,\rangle$, and hence $\rho^*_M(\Delta^*_M(\langle\,\rangle)) = \langle\,\rangle$.

- If $x = (k, d)\,\&\,y$, then for an arbitrary $M$:

$$\begin{aligned}
&\rho^*_M(\Delta^*_M((k,d)\,\&\,y)) = \\
&\quad \rho^*_M((k, \Delta(M(k), d))\,\&\,\Delta^*_{M[k \mapsto d]}(y)) = \\
&\quad (k, \rho(M(k), \Delta(M(k), d)))\,\&\,\rho^*_{M[k \mapsto \rho(M(k), \Delta(M(k),d))]}(\Delta^*_{M[k \mapsto d]}(y)) = \\
&\quad (k, d)\,\&\,\rho^*_{M[k \mapsto d]}(\Delta^*_{M[k \mapsto d]}(y)) = \\
&\quad (k, d)\,\&\,y
\end{aligned}$$

To prove the second premise of the behavioral refinement rule with invariant is then straightforward.

The structure of the system remains unchanged.

**Step 7: Removing an input channel.** Since the behavior of RDB now depends only on the data on R, and not on that in I, we can disconnect I from RDB. The channel I now only feeds the encoder.

The new system looks as follows:

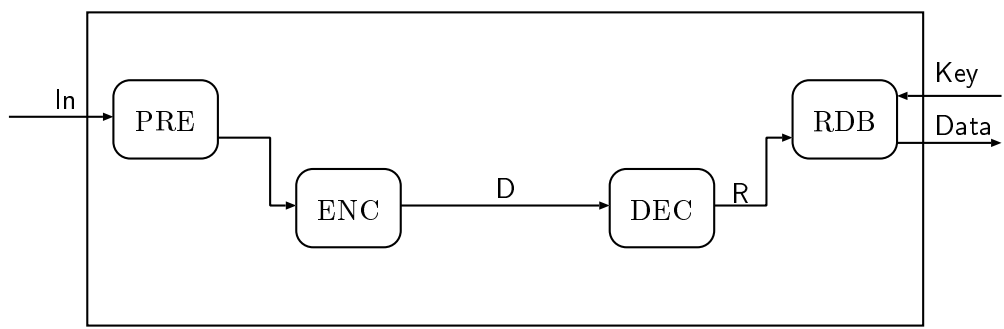

**Step 8: Folding subsystems.** In the last refinement step, we fold the two components PRE and ENC to a new component PRE′, and DEC together with RDB to a new component RDB′:

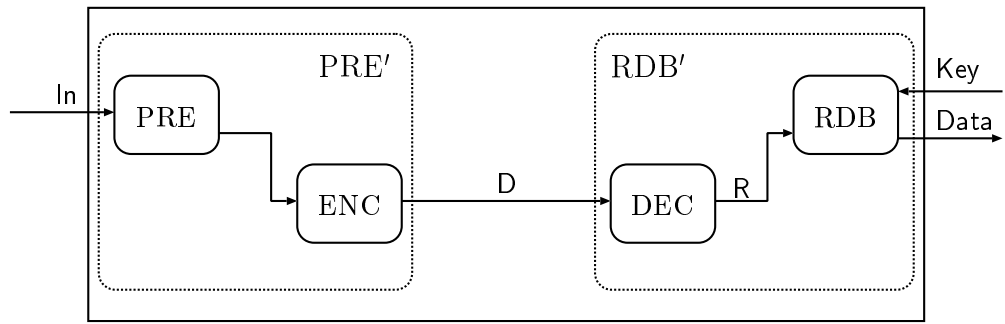

**Comments on the transformation.** The refinement steps described above are not fully formal; they cannot be, since we did not use a properly formalized description of the component behaviors. Of course, state transition diagrams can be given a mathematical semantics [5], and in [11] a refinement calculus for state transition diagrams is defined. We hope, however, that the example shows that although each individual refinement rule is quite simple, they can be used together for complex system transformations.

As expected, the behavioral rule with invariant is the most complex rule to apply. In general, it is a difficult task for the system designer to find a proper invariant $\Psi$ that is both easy to establish and sufficiently strong to use. The maximal invariant $\Psi(l) = \mathit{True}$ leads to our initially given simple refinement rule without an invariant. The minimal possible $\Psi$ gives an exact description of the internal behavior of a system, but it is often difficult to find and too complex to use.

If one wants to change the behavioral descriptions of a component — in our example, to change the remote database so that it stores the data on R instead of that on I —, one can take advantage of his knowlege about the dependencies between internal streams. Thus, when designing systems for adaptility, one should strive not for efficiency, but first for clarity of design, where information in all channels is as explicit as possible. Later, refinement steps should remove the redundancy to gain an efficient implementation of the system.

# 6 Conclusion

We believe that the question of how to manipulate and adapt an architecture during system development has not been adequately addressed so far. In particular, a basic calculus, dealing with simple addition ond removing of channels and components in an architectural style has—to our knowledge—not been considered before.

The most promising attempt at architecture refinement so far has been given in [8, 9]. In that work, data flow architectures are implemented by shared-memory architectures. However, the semantics used is not particularly well-suited for data flow, and they do not seem to support nondeterminism or underspecification. Hence they only allow "faithful implementation" which is in contrast to our approach. They do not allow adding or removing data flow connections, which seems to stem from the lack of support for underspecification in their model. Underspecification is the primary source that allows us to change the information structure of an architecture. In our history-based semantics, underspecification can be easily handled.

We think that a simple refinement calculus, especially one well-suited for the graphical manipulation of data flow networks, is crucial for the applicability of a formal method. The calculus defined in this paper allows to reuse given architectures or architectural patterns and to adapt them to specific needs. It is therefore interesting to develop a library of dataflow architecture designs for different applications.

Our calculus currently only deals with refinement internal to the system. As future work, we will extend it with rules to change the interface signature in the style of [1]. The new rules will allow us to change the input or output channels of a system, as well as to split one channel into several channels carrying parts of the original information or vice versa.

Another interesting direction is the description of component behaviors by state machines and the application of state machine refinement rules (as defined e.g. in [11]) for component behavior refinement. A concrete description technique for the component behaviors is essential for the proof of the invariant in the behavior refinement rule. We have defined our calculus so that it can be incorporated into CASE tools. A prototypical tool, AUTOFOCUS [7], is currently under development at our department. AUTOFOCUS already has a graphical syntax for system structures similar to the ones we use, and also provides state-machine-based specification mechanisms for component behavior.

Finally, architecture refinement is by no means limited to business information systems. Another promising application area is hardware design and in particular the codesign of hardware and software components, where frequently a basic design has be changed because of cost or performance considerations. Moreover, the simpler description techniques used in hardware design and the finite-state nature of such systems open the door to automatic verification of the refinement rule premises.